\documentclass[apj]{emulateapj}

\received{June 25 2009}
\accepted{July 1 2009}

\def\simgt{\lower.5ex\hbox{$\; \buildrel > \over \sim \;$}}
\def\simlt{\lower.5ex\hbox{$\; \buildrel < \over \sim \;$}}

\def\etal{{et~al.}}
\def\amin{\ifmmode^{\prime}\else$^{\prime}$\fi}
\def\asec{\ifmmode^{\prime\prime}\else$^{\prime\prime}$\fi}

\def\simgt{\lower.5ex\hbox{$\; \buildrel > \over \sim \;$}}
\def\simlt{\lower.5ex\hbox{$\; \buildrel < \over \sim \;$}}

\newcommand\chandra{{\it Chandra}}

\newcommand\xmm{{\it XMM-Newton}}

\newcommand\hess{{HESS}}

\newcommand\integral{{\it INTEGRAL\/}}
\newcommand\INTEGRAL{{\it INTEGRAL\/}}

\newcommand\fermi{{\it Fermi\/}}

\newcommand\snr{G12.82$-$0.02}
\newcommand\tev{\hbox{HESS~J1813$-$178}}
\newcommand\igr{\hbox{IGR~J18135--1751}}
\newcommand\mev{\hbox{0FGL~J1814.3--1739}}
\newcommand\cxo{CXOU J181335.1$-$174957}
\newcommand\psr{PSR~J1813$-$1749}

\slugcomment{Submitted to The Astrophysical Journal Letters}

\begin{document}

\title{Discovery of a Highly Energetic X-ray Pulsar Powering \tev\
in the Young Supernova Remnant \snr} 

\author{E. V. Gotthelf \&  J. P. Halpern}

\affil{Columbia Astrophysics Laboratory, Columbia University, 550 West
120$^{th}$ Street, New York, NY 10027, USA}

\begin{abstract}

We report the discovery of 44.7~ms pulsations from the X-ray source
\cxo\ using data obtained with the \xmm\ Observatory.  \psr\ lies near
the center of the young radio supernova remnant \snr, which overlaps
the compact TeV source \tev. This rotation-powered pulsar is the
second most energetic in the Galaxy, with a spin-down luminosity of
$\dot E = (6.8\pm 2.7) \times 10^{37}$~erg~s$^{-1}$. In the rotating
dipole model, the surface dipole magnetic field strength is $B_s =
(2.7\pm0.6) \times10^{12}$~G and the spin-down age $\tau_c \equiv
P/2\dot P = 3.3-7.5$~kyr, consistent with the location in the
small, shell-type radio remnant.
At an assumed distance of
4.7~kpc by association with an adjacent young stellar cluster, the efficiency
of \psr\ in converting spin-down luminosity to radiation is $\approx 0.03\%$
for its 2$-$10~keV flux, $\approx 0.1\%$ for its $20-100$~keV
\integral\ flux,
and $\approx 0.07\%$ for the $>200$ GeV emission of \tev, making it a
likely power source for the latter.  The nearby young stellar
cluster is possibly the birthplace of the pulsar progenitor, as well
as an additional source of seed photons for inverse Compton scattering
to TeV energies.

\end{abstract}

\section{Introduction}

The identification of high-energy (TeV) $\gamma$-ray emission in our
Galaxy with supernova products has opened up new prospects for
understanding the energetics of these stellar remnants.
Over half of the $>50$ Galactic TeV sources are identified with 
supernova remnants (SNRs) or pulsar
wind nebulae (PWNe)\footnote{VHE $\gamma$-ray Sky Map and Source Catalog,
http://www.mppmu.mpg.de/\~\,rwagner/sources/}.
Several physical mechanisms are available for generating the observed
$\gamma$-ray photons, including non-thermal bremsstrahlung, and
inverse-Compton scattering of ambient microwave, IR, or optical
photons off relativistic electrons. $\gamma$-rays can also be produced
in hadronic collisions of high-energy protons off local material, from
the decay of their neutral pion products. An unambiguous detection of
the latter process associated with a Galactic SNR would provide direct
confirmation of cosmic-ray production in SNR shocks.

A unique opportunity to explore these high energy mechanisms is
provided by the TeV source \tev, one of the brightest and most compact
objects located by the \hess\ Galactic Plane Survey \citep{aha05a}.
Within the TeV extent lies \snr, a previously uncatalogued young
shell-type radio supernova remnant
\citep{bro05,hel05}. Furthermore, there is evidence for
broad-band high-energy emission.  Overlapping the SNR shell lies a
2--10~keV X-ray source AX~J1813--178 \citep{bro05,hel05}, the \integral\
soft gamma-ray (20-100~keV) source \igr\ \citep{ube05}, and
possibly GeV emission from the \fermi\
source \mev\ \citep{abd09}. Follow-up high resolution X-ray
studies resolved the X-ray emission into a point source and bright
surrounding nebula, whose properties indicate that a young, energetic
rotation-powered pulsar \citep{fun07b,hel07} is responsible for the
extended X-ray emission and, ultimately, the broad-band high-energy
radiation.  A deep radio timing search failed to detect pulsations
\citep{hel07}.

In this Letter, we report on a long, continuous \xmm\ X-ray timing
observation of \cxo, the central point source in \snr,
resulting in the discovery of \psr. We show that the spin-down
luminosity of the pulsar is sufficient to power the broad-band X-ray
and $\gamma$-ray emission.
A nearby young stellar cluster \citep{mes09}
may provide seed photons for upscattering to $\gamma$-rays,
and is a possible birth place of the pulsar progenitor.

\section{\xmm\ Observations and Results}

The \chandra\ source \cxo\ was observed by \xmm\ with a 98~ks exposure
on 2009 March 27 using the European Photon Imaging Camera (EPIC;
\citealt{tur03}). The EPIC pn CCD was operated in small-window mode ($4\farcm3 \times
4\farcm3$ field-of-view; 29\% deadtime). This mode provides 5.7~ms
time resolution, allowing a search for even the most rapidly rotating
young pulsar. \xmm\ is sensitive to X-rays in the
nominal $0.1-12$~keV range with energy resolution $\Delta E/E \approx
0.1/\sqrt{E({\rm keV})}$. The target was placed at the default EPIC~pn
focal plane location for a point source.  Data was also
acquired with the two EPIC~MOS CCD cameras (MOS1 and MOS2). These 
were operated in full frame mode with a time resolution of
2.6~s, insufficient to search for a typical pulsar signal.
The medium filter was used for both EPIC instruments.

All data were processed using the SAS version xmmsas\_20060628\_1801-7.0.0
pipeline, and were analyzed using both the SAS and FTOOLS software
packages. The observation was free of significant particle
contamination and provided a near continuous 97.9~ks of good observing time
per detector. Photon arrival times from the pn CCD
were transformed to the solar system barycenter in Barycentric
Dynamical Time (TDB) using the JPL DE200 ephemeris and the \chandra\
measured coordinates given in \cite{hel07}.

To search for the expected pulsed signal in the pn CCD, $2-10$~keV photons were
extracted from an 0\farcm5 diameter aperture centered on the source,
optimized for the highly absorbed point-source embedded in the
substantial nebular emission (see \citealt{hel07}). A $2^{25}$-bin fast
Fourier transform (FFT) search algorithm was used and a highly
significant signal detected at $P = 44.7$~ms.  We performed a refined
$Z^2_1$ (Rayleigh) search \citep{buc83} centered on the FFT signal and
localized the pulsed emission with peak power $Z^2_1 =
476$. This corresponds to essentially zero false detection
probability for the number of independent search trials. The pulse
profile is slighly narrower then a sinusoid and shows no
energy dependence in subdivided bands within the 2$-$10~keV range.

\begin{figure}[t]
\hspace{-0.1in}
\centerline{
\hfill
\includegraphics[height=0.9\linewidth,angle=270,clip=true]{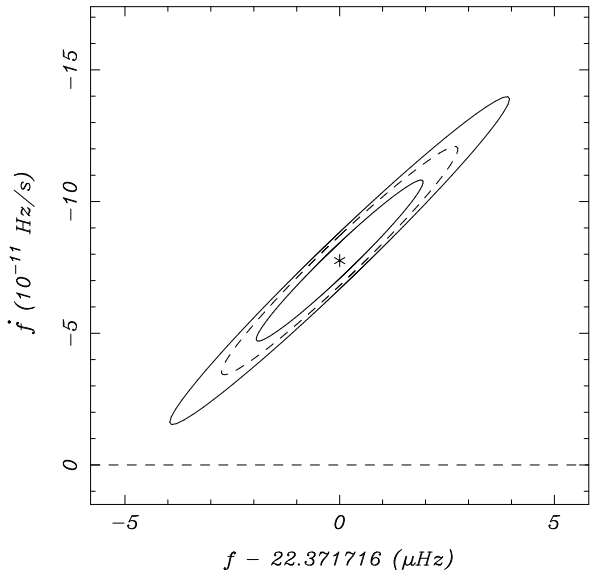}
\hfill
}
\caption{Discovery of \psr\ using \xmm\ EPIC~pn data of 2009 March
27. Contours of the signal power, realized by a $Z^2_1$ search over
frequency and frequency derivative ($f,\dot f$), corresponding to the
68\%, 90\%, and 95\% confidence levels for two interesting parameters.}
\label{ellipse}
\end{figure}

Given the long time span of the data, we also looked for the effect of a
period derivative $\dot f$ on the pulsar signal by constructing a
$Z^2_1$ search over ($f,\dot f$)-space.  Figure~\ref{ellipse}
shows the resulting contours for the signal power at the 68\%, 90\%,
and 95\% confidence level for two interesting parameters. The formal
result is $f = 22.371716(2)$~Hz and $\dot f = -(7.7\pm 3.1)\times
10^{-11}$~Hz~s$^{-1}$; the uncertainties are $1\sigma$.   This large
$\dot f$ is detectable in the single long
observation, as the quadratic term ${1 \over 2}\dot f (t-t_0)^2$
in the phase ephemeris contributes $-0.37$ cycles of
rotation over the elapsed time.  However, its value is
highly uncertain at the 95\% confidence level, as shown
in Figure~\ref{ellipse}.  A precise value of $\dot f$
will be determined easily with a follow-up observation.
The best value and $1 \sigma$ range for the spin-down parameters
in the dipole pulsar model are spin-down power $\dot E \equiv
-4\pi^2 I f \dot f = (6.8\pm 2.7) \times 10^{37}$~erg~s$^{-1}$,
surface dipole magnetic field strength $B_s \equiv 3.2 \times 10^{19}
\sqrt{-f\dot/ f^3} = (2.7 \pm 0.6) \times 10^{12}$~G, and characteristic
age $\tau_c \equiv -f/2\dot f = 3.3-7.5$~kyr. The pulsar parameters
are listed in Table~\ref{ephem}.

\begin{deluxetable}{ll}
\tabletypesize{\small}
\tablecaption{\label{ephem}Timing Parameters of \psr }
\tablehead{
\colhead{Parameter}   &
\colhead{Value}   }
\startdata                                       
R.A. (J2000)\tablenotemark{a}\dotfill          & $18^{\rm h}13^{\rm m}35.166^{\rm s}$\\
Decl. (J2000)\tablenotemark{a}\dotfill         & $-17\arcdeg49'57.48^{\prime\prime}$    \\
Epoch (MJD)\dotfill                            & 54917                \\
Period, $P$ (ms)\dotfill                       & 44.699297(4)              \\
Period derivative, $\dot P$\dotfill            & $(1.5 \pm 0.6)\times10^{-13}$       \\
Characteristic age, $\tau_c$ (kyr)\dotfill      & 3.3-7.5                       \\
Spin-down luminosity, $\dot E$ (erg\,s$^{-1}$)\dotfill & $(6.8 \pm 2.7)\times10^{37}$   \\
Surface dipole magnetic field, $B_s$ (G)\dotfill & $(2.7\pm0.6)\times10^{12}$
\enddata
\tablecomments{\footnotesize $1\sigma$ uncertainties are given.}
\tablenotetext{a}{\footnotesize Chandra ACIS-I position from \cite{hel07}.}
\end{deluxetable}

\begin{figure}[b]
\centerline{
\hfill
\includegraphics[height=0.9\linewidth,angle=270,clip=true]{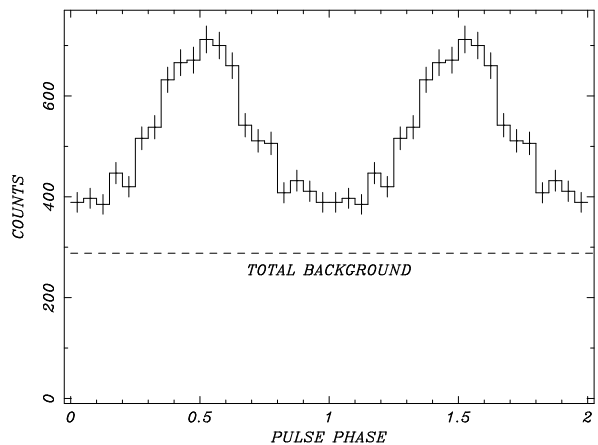}
\hfill
}
\caption{Pulse profile of \psr\ folded in 20 phase bins at the
frequency corresponding to the peak power in the ($f,\dot f$) search
periodogram shown in Figure~\ref{ellipse}. The dashed line indicates
the combined contributions from the X-ray background, PWN, and
detector contamination. Phase zero is arbitrary.}
\label{profile}
\end{figure}

Figure~\ref{profile} presents the pulse profile of \psr\
corresponding to the data folded at the peak signal value in
($f,\dot f$)-space.  The detected pulsed fraction for the extracted
photons is $25\pm3\%$, defined here as $N({\rm
pulsed})/N({\rm total})$, where we choose the minimum of the folded
light curves as the unpulsed level. The quoted uncertainties are
derived by propagating the counting statistics of the light curve for
10 bins.

\begin{figure}[t]
\centerline{
\hfill
\includegraphics[height=0.9\linewidth,angle=270,clip=true]{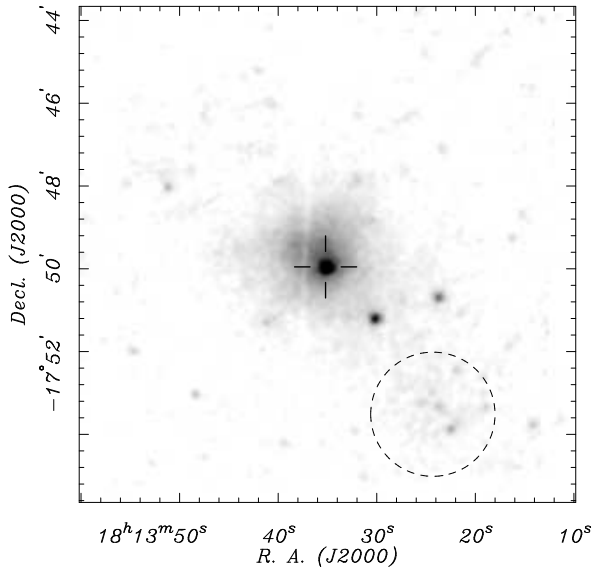}
\hfill
}
\caption{\xmm\ EPIC~MOS broad-band X-ray image of
\psr\ and its wind nebula. The image is exposure corrected, smoothed,
and scaled to highlight the PWN emission and faint sources; the
pulsar ({\it cross}) is saturated in this display. 
Sources associated with the core of
the young stellar cluster of \citet{mes09} are seen to the southwest
({\it dashed circle}).}
\label{fig:mosimage}
\end{figure}

The intrinsic pulsed fraction is difficult to determine due to the
background contamination from the PWN in the source aperture (see
Figure~\ref{fig:mosimage}).  Using the high-resolution \chandra\ X-ray
image of \snr\ \citep{hel07}, we can resolve the pulsar emission from
that of the nebula for a given sized aperture to estimate their
relative contributions. Over the EPIC~pn signal extraction aperture,
allowing for the EPIC~pn beam size, we derive a photon flux ratio of
$F_{\rm PWN}/F_{\rm PSR+PWN} = 0.51$, after correcting for the
\chandra\ background ($\approx 15\%$ of PWN counts). Assuming a
similar relative spectral response between the two detectors, we use
the \chandra\ ratio to correct the background-subtracted XMM EPIC~pn
counts to determine the PWN contamination in the source aperture.
After correcting for the EPIC~pn X-ray and detector background
contributions, and taking into account the \chandra\ derived PWN
contamination, we determine an intrinsic pulsed fraction for \psr\ of
$58\pm9\%$.

\section{Discussion}

\subsection{Age and Energetics}

HESS~J1813$-$178 is one of the more compact TeV sources
to be associated with a
SNR.  The Gaussian extent of the source is only
$\sigma = 2.2^{\prime} \pm 0.4^{\prime}$ \citep{aha06a},
while the radio shell of \snr\ is only
$2.5^{\prime}$ in diameter \citep{bro05}.
It is not clear if these are significantly different.
Unlike \psr/\tev,
many TeV sources associated with pulsars show a
notable displacement between the two, e.g., PSR B0833$-$45/Vela~X
\citep{aha06b},
PSR B1823$-$13/HESS J1825$-$137 \citep{aha06c,pav08}, PSR
J1809$-$1917/HESS J1809$-$193 \citep{aha07,kar07}, PSR
J1718$-$3825/HESS J1718$-$385 \citep{aha07,hin07}, and possibly PSR
J1617$-$5055/HESS J1616$-$508 \citep{aha06a,lan07}. For these
systems, the TeV source is larger than the X-ray PWN.
Apparently, the electrons responsible for TeV emission via inverse-Compton
scattering have longer
lifetimes, and may have been injected at earlier times,
than the X-ray synchrotron emitting electrons now in the inner
PWN.  It is also possible that the PWN is displaced in one direction
by an asymmetric reverse shock resulting from a supernova that
exploded initially in an inhomogeneous medium \citep{blo01}.
These systems contain ``middle-aged'' pulsars, $>10^4$~yr, 
which are old enough for the reverse shock to have had its effect,
or for relativistic electrons to have diffused out of the high $B$-field
PWN.  Theoretically, a TeV source powered by a
pulsar can live for up to 100~kyr as an inverse Compton emitter
in regions of low ($\mu$G) $B$-fields, with luminosity even exceeding
the current spin-down power of the pulsar \citep{dej08}.
Their TeV luminosities can then exceed their X-ray luminosities
by a large factor, which is observed for several sources.

\begin{deluxetable*}{lrllll}
\tabletypesize{\small}
\tablecaption{\label{flux}Multi-waveband Luminosity Measurements of \psr }
\tablehead{
\colhead{Mission} & \colhead{Energy Band}& \colhead{Source} & \colhead{Luminosity\tablenotemark{a}}      & \colhead{$\epsilon$\tablenotemark{b}} & \colhead{Reference} \\
\colhead{}        & \colhead{}           & \colhead{}            & \colhead{(erg~s$^{-1}$)} & \colhead{(\%)}   & \colhead{}
}					 					 
\startdata                                                          
    \chandra\     &   2--10~keV & AX~J1813--178 & $1.9 \times 10^{34}$ & 0.03 & \cite{hel07} \\
   \INTEGRAL\     & 20--100~keV & \igr\         & $7.3 \times 10^{34}$ & 0.1  & \cite{dea08} \\
      \fermi\tablenotemark{c}     &  0.1--1~GeV & \mev\         & $5.4 \times 10^{35}$ & 0.8  & \cite{abd09} \\
       \hess\     &  $>200$~GeV & \tev\         & $4.8 \times 10^{34}$ & 0.07 & \cite{aha06a}
\enddata
\tablenotetext{a}{\footnotesize Total luminosity in the band assuming $d=4.7$~kpc.} 
\tablenotetext{b}{\footnotesize Efficiency in the band assuming spin-down power
$\dot E = 6.8 \times 10^{37}$ erg~s$^{-1}$.}
\tablenotetext{c}{\footnotesize Conversion from \fermi\ photon flux assumes
a power-law photon index $\Gamma=2$.  However, association of the \fermi\
source with \snr/\psr\ is not yet established (see \S 3.2).}
\end{deluxetable*}

In the case of \psr/\tev, the compact TeV source and lack of relative
offset can now be explained by its youth, like the Crab and several
young pulsars with compact \hess\ sources that are co-located with
their X-ray PWNe. The other compact sources are PSR J1846$-$0258/HESS
J1846$-$029 in the SNR Kes~75 \citep{dja08}, PSR J1833$-$1036/HESS
J1833$-$105 in G21.5$-$0.9 \citep{dja08}, and the newly discovered
52~ms PSR J1747$-$2809/HESS J1747$-$281 in G0.9+0.1
\citep{cam09,aha05b}.  PSR~J1846$-$0258 is a $0.325$~s pulsar with a
characteristic age $\tau_c$ of only 728~yr, and a spin-down luminosity
$\dot E = 8 \times 10^{36}$ erg~s$^{-1}$ \citep{got00}.  PSR
J1833$-$1036 has $\tau_c = 4900$~yr and $\dot E = 3.3 \times 10^{37}$
erg~s$^{-1}$.  However, the actual age of PSR J1833$-$1036 is
probably only $\sim 870$~yr as measured by the expansion of
G21.5$-$0.9 \citep{bie08}.  PSR J1747$-$2809 has $\dot E = 4.3 \times
10^{37}$ erg~s$^{-1}$ and $\tau_c = 5300$~yr.  \psr\ shares with
these young pulsars a characteristic age $\simlt5000$~yr, but its true
age may be significantly smaller than this because its initial period
$P_0$ is not necessarily $<<P$.  In fact, \cite{bro05} estimate an age
of $<2500$~yr for \snr\ based on its size, and possibly as small as
300~yr if it is still in the free-expansion stage.  In these
relatively young and compact PWNe the high $B$-fields make for
efficient synchrotron X-ray sources, and inefficient inverse-Compton
TeV emission.  The Crab, PSR J1833$-$1036, and PSR J1846$-$0258, have
ratios $L_x/L_{\gamma}$ of $\approx 120$, 40, and 10, respectively
\citep{dja08}.  This ratio is $\approx 0.5$ for \psr.

The efficiency of \psr\ converting spin-down luminosity to radiation at
a distance of $d=4.7$~kpc is $\approx 0.03\%$ for the 2$-$10~keV X-ray flux,
$\approx 0.1\%$ for the $20-100$~keV \integral\ flux,
and $\approx 0.07\%$ for the $>200$ GeV emission of \tev\ (see Table~\ref{flux}).
\citet{mat09} used empirical correlations of the ratio
$L_x/L_{\gamma}$ with $\dot E$
among known \hess\ PWNe to estimate that
$\dot E \sim 1.5 \times 10^{37}$ erg~s$^{-1}$
and $\tau_c \sim 6$~kyr for the as-yet undiscovered \psr,
which turns out to be a reasonably accurate prediction
of their actual values.

\subsection{Environment and Associations}

The location of \snr\ near a young stellar cluster rich in massive
binaries and containing a second supernova remnant (G12.72--0.00)
suggests an association \citep{mes09}. 
Those authors determined cluster membership for giants and
early-type stars, and derived
a consistent distance of 4.7~kpc using radial
velocities and optical/infrared extinction.
They estimated an age of 6--8~Myr, and a total initial cluster
mass of $2000-6500 M_{\odot}$.   It is
plausible that the progenitors of both \snr\ and G12.72--0.00
were born in the cluster, and had masses similar to those
of the red supergiant and WR stars now present, $20-30 M_{\odot}$.
The cluster is centered $4.4^{\prime}$ southwest of \snr\
(see Figure~\ref{fig:mosimage}).  At this offset, a 5~kyr
old neutron star born in the center of the cluster requires a
transverse velocity of $\approx 1200$~km~s$^{-1}$ to reach its present
location, or less if it was born in the outskirts close to its
present location.

It is plausible that seed IR/optical photons from this star cluster
enhance TeV emission from \tev, via inverse
Compton scattering in its extended PWN, above the
minimum expected from the cosmic microwave background.
Such a role was also hypothesized by \citet{hel07} for the adjacent
W33 star-forming region, at an estimated distance of 4.3~kpc.
Although these objects are not necessarily all at the same distance,
they could be.

Pulsed emission from \psr\ is expected to be detectable by \fermi, and
the GeV band could be where the luminosity peaks, as it does in many
energetic pulsars.  Although a possible association of the \fermi\
source \mev\ with \tev/ \snr\ has been noted by \citet{abd09}, the
SNR lies just outside the large, $11^{\prime}$ radius 95\%
confidence \fermi\ error circle.  It is not yet possible to
identify definitively the GeV emission with the pulsar,
which would require detection of pulsed $\gamma$-rays.
The implied luminosity of \mev\ (assumed isotropic) in Table~\ref{flux}
is consistent with the trend of efficiencies
among $\gamma$-ray pulsars \citep{tho04}.
Accumulation of more exposure with \fermi\ and detailed 
investigation of this region could be informative.

\section{Conclusions}

Using \xmm, we discovered \psr, a highly energetic 44.7~ms pulsar
associated with the supernova remnant \snr\ and powering its PWN
from X-ray to TeV energies.
Its preliminary spin-down properties make it one of the most
energetic pulsars in the Galaxy, possibly second only to the Crab.
Its apparent youth explains the small extent of its
SNR and TeV nebula relative to older pulsars.
The high $\dot E$ and young age support a
one-zone synchrotron/IC model of the type 
applied by \citet{fun07b}, in which \psr\
accelerates electrons to $>10^{15}$~eV. 
The lack of an X-ray SNR shell, coupled with a high spin-down power for
\psr, argues that the PWN, not the SNR,
is almost certainly the TeV source.

\acknowledgements

This investigation is based on observations obtained with \xmm, an ESA
science mission with instruments and contributions directly funded by
ESA Member States, and NASA.  Support for this work was provided by
NASA through \xmm\ grant NNX08AJ45G.  E.V.G thanks Dr. Djannati-Ata\"i
for hospitality and support during his visit to APC, Universiti{\'e}
Paris-Diderot.

\end{document}